\documentstyle[aps,twocolumn]{revtex}
\begin{document}
\draft
\title{Correlations of Eigenfunctions in Disordered Systems}
\author{Ya.~M.~Blanter$^{a,b}$ and A.~D.~Mirlin$^{a,c}$}
\address{$^a$ Institut f\"ur Theorie der Kondensierten Materie,
Universit\"at Karlsruhe, 76128 Karlsruhe, Germany\\
$^b$ Department of Theoretical Physics, Moscow Institute for Steel and
Alloys, Leninskii Pr. 4, 117936 Moscow, Russia\\
$^c$ Petersburg Nuclear Physics Institute, 188350 Gatchina,
St. Petersburg, Russia}
\date{\today}
\maketitle 
\tighten
\begin{abstract}
Correlations of  eigenfunctions, 
$\langle|\psi_k(r_1)|^2|\psi_l(r_2)|^2\rangle$, in a disordered system are
investigated. We derive general formulae expressing these
correlation functions in terms of the supermatrix sigma-model. In
particular case of weak localization regime we find that the
correlations of the same eigenfunction are proportional to $g^{-1}$
for large distances, while the correlations of two different
eigenfunctions cross over from $g^{-1}$ behavior for $r_1=r_2$ to
$g^{-2}$ one for $\vert r_1 - r_2 \vert \gg l$, with $g$ and $l$ being the
dimensionless conductance and the mean free path, respectively. 
\end{abstract}

Statistics of eigenfunction fluctuations in disordered and chaotic
systems have attracted a research interest recently. The fluctuations
of eigenfunction amplitudes determine statistical properties of
conductance peaks and level width in quantum dots in the Coulomb
blockade regime \cite{Jal,PEI,MPA,Chang,Folk}, and can be directly
measured in the microwave cavity experiments
\cite{Stockmann,Sridhar}. On the theoretical side, the recent progress
is based on application of the supersymmetry method to the problem of
eigenfunction statistics \cite{MF1,PEI}. It was found
that the distribution of eigenfunction amplitudes is in the leading
(zero-mode) approximation correctly described by formulae of the
random matrix theory (RMT). Deviations from the RMT predictions were
studied in Refs. \cite{MF1,MF,tails}. Correlations of amplitudes of an
eigenfunction in two different spatial points were considered in
Ref. \cite{Prig} on the level of zero-mode approximation; the
latter was shown \cite{Srednicki} to be equivalent to the RMT-like
assumption of the Gaussian fluctuations of wavefunctions. 

All results mentioned above concern fluctuations of the {\em same}
eigenfunction. In the present Letter we study for the first time
correlations of 
amplitudes of two {\em different} eigenfunctions. We derive
general expressions in terms of the supermatrix sigma model, 
valid for arbitrary diffusive (or classically chaotic) system, and then
apply them to the weak localization regime.

In order to evaluate the correlations of the wavefunctions we use the
technique similar to that of Ref. \cite{GErev}. Namely, we consider a
quantity
\begin{eqnarray} \label{a1}
& & A(\bbox{ r}_1, \bbox{ r}_2, \omega) \\ &&= 
\left\langle \sum_{k,l} \vert \psi_k(\bbox{ r}_1) \psi_l(\bbox{
r}_2) \vert^2 \delta (\epsilon - \epsilon_k) 
 \delta(\epsilon + \omega
- \epsilon_l) \right\rangle - \nonumber\\ &&\left\langle \sum_{k} \vert
\psi_k(\bbox{ r}_1) \vert^2 \delta(\epsilon - \epsilon_k)
\right\rangle \left\langle \sum_l
\vert \psi_l(\bbox{ r}_2) \vert^2 \delta(\epsilon + \omega
- \epsilon_l) \right\rangle . \nonumber 
\end{eqnarray}
Here the angular brackets denote the impurity average. We have
introduced the eigenstates $\psi_k (\bbox{
r})$ and eigenvalues $\epsilon_k$ of the Hamiltonian $\hat H = \hat
H_0 + U(\bbox{r})$ in a particular
disorder configuration $U(\bbox{\bf r})$, $H_0$ being the Hamiltonian
of the free particle. Further we denote 
\begin{eqnarray} \label{fun}
\alpha(\bbox{ r}_1, \bbox{ r}_2, \epsilon_k) & = & \left\langle \vert
\psi_k(\bbox{ r}_1) \psi_k(\bbox{ r}_2) \vert^2 \right\rangle
\nonumber \\ 
\beta(\bbox{ r}_1, \bbox{ r}_2, \epsilon_k, \epsilon_l) & = &
\left\langle \vert \psi_k(\bbox{ r}_1) \psi_l (\bbox{ r}_2)
\vert^2 \right\rangle, \ \ \ k \ne l,
\end{eqnarray}
where the averaging is carried over all eigenstates $\psi_k, \psi_l$
with given values of the energies $\epsilon_k, \epsilon_l$. Since we
are interested in a relatively narrow window in the spectrum, $\omega
\ll \epsilon_F$, these quantities are translationally invariant:
\begin{eqnarray} \label{transinv}
\alpha(\bbox{ r}_1, \bbox{ r}_2, \epsilon_k) & = & \alpha(\bbox{ r}_1,
\bbox{ r}_2), \nonumber \\  
\beta(\bbox{ r}_1, \bbox{ r}_2, \epsilon_k, \epsilon_l) & = &
\beta(\bbox{ r}_1, \bbox{ r}_2, \epsilon_k - \epsilon_l).
\end{eqnarray}
Due to the presence of delta-functions in Eq. (\ref{a1}), we can
rewrite the latter {\em exactly} as
\begin{eqnarray} \label{a3}
A(\bbox{r}_1,\bbox{r}_2,\omega) & = & \alpha (\bbox{r}_1, \bbox{r}_2)
\Delta^{-1} \delta(\omega) \nonumber \\
& + & \beta(\bbox{r}_1,\bbox{r}_2,\omega)
\Delta^{-2} R(\omega), 
\end{eqnarray}
where $\Delta$ is the mean level spacing, $\Delta = (\nu V)^{-1}$,
with $V$ and $\nu$ being the system volume and the density of states,
respectively. Here we have introduced the two-level correlation
function, 
\begin{equation} \label{corf}
R(\omega) = \Delta^2 \left\langle \sum_{k \ne l} \delta(\epsilon -
\epsilon_k) \delta(\epsilon + \omega - \epsilon_l) \right\rangle .
\end{equation}
We would like to stress that all transformations up to now are
completely rigorous. In particular, we {\em do not} assume any
decoupling of eigenfunction and eigenvalue correlations.

On the other hand, the quantity (\ref{a1}) can be written in terms
of the Green's functions in the coordinate-frequency representation,
\begin{eqnarray} \label{first}
& & A(\bbox{ r}_1, \bbox{ r}_2, \omega) = (2\pi^2)^{-1}
\nonumber \\
\times & & \mbox{Re}
\left\{ \langle G^R(\bbox{ r}_1, \bbox{ r}_1, \epsilon) G^A
(\bbox{ r}_2, \bbox{ r}_2, \epsilon + \omega) \rangle \right.
\nonumber \\
& & - \left. \langle G^R(\bbox{ r}_1, \bbox{ r}_1, \epsilon) \rangle
\langle G^A (\bbox{ r}_2, \bbox{ r}_2, \epsilon + \omega)
\rangle \right\}.
\end{eqnarray}
The expression (\ref{first}) can be directly
calculated with the use of the supersymmetry technique 
\cite{Efetov,Verb,Zirn}. We concentrate in the sequel on the case of broken
time-reversal symmetry (unitary ensemble); generalization to the other
ensembles is straightforward. After the standard manipulations we get  
\begin{eqnarray} \label{Q1}
& & A (\bbox{ r}_1, \bbox{ r}_2, \omega) = - (2\pi^2)^{-1}
\mbox{Re} \left\{ \left\langle g_{bb}^{11} (\bbox{ r}_1, \bbox{
r}_1) g_{bb}^{22} (\bbox{ r}_2, \bbox{ r}_2) 
\right. \right. \nonumber \\ 
& & + \left. \left. g_{bb}^{12} (\bbox{ r}_1, \bbox{
r}_2) g_{bb}^{21} (\bbox{ r}_2, \bbox{ r}_1) \right\rangle_F  -
\left\langle g_{bb}^{11} (\bbox{ r}_1, \bbox{
r}_1) \right\rangle_F \right. \nonumber \\
& & \left. \times \left\langle g_{bb}^{22} (\bbox{ r}_2,
\bbox{ r}_2) \right\rangle_F \right\}.
\end{eqnarray}
Here $\langle \dots \rangle_F$ denotes the averaging with the action
of the supermatrix sigma-model $F[Q]$:
\begin{eqnarray} \label{aver1}
& &\langle \dots \rangle_F = \int DQ ( \dots ) \exp(-F[Q]), \nonumber \\ 
& & F[Q] = - \frac{\pi\nu}{4} \int d\bbox{ r} \ \mbox{Str} [D(\nabla
Q)^2 + 2i(\omega + i0) \Lambda Q],
\end{eqnarray}
where $D$ is the diffusion coefficient, 
$Q = T^{-1}\Lambda T$ is a 4$\times$4 supermatrix, $\Lambda =
\mbox{diag} (1,1,-1,-1)$, and $T$ belongs to the supercoset space
$U(1,1 \vert 2)/U(1\vert 1)\times U(1 \vert 1)$. The symbol
$\mbox{Str}$ denotes the supertrace defined as 
$\mbox{Str} B = B_{bb}^{11} - B_{ff}^{11} + B_{bb}^{22} - B_{ff}^{22}$. The 
upper matrix indices correspond to the retarded-advanced
decomposition, while the lower indices denote the boson-fermion
one. The Green's function $g$ in Eq. (\ref{Q1}) is the solution to
the matrix equation:
\begin{eqnarray} \label{Green}
& & \left[ -i(\epsilon + \frac{\omega}{2} - \hat H_0) -
\frac{i}{2}(\omega + i0)\Lambda + Q/2\tau \right] g(\bbox{
r}_1, \bbox{ r}_2) \nonumber \\
& & = \delta (\bbox{ r}_1 - \bbox{ r}_2).
\end{eqnarray}
Expressing these functions through the matrices $Q$ and taking into
account Eq. (\ref{a3}), we arrive at the following equation valid in arbitrary
diffusive system:
\begin{eqnarray} \label{genex}
& & 2\pi^2 \left[ \frac{\alpha(\bbox{r}_1, \bbox{r}_2)}{\Delta}
\delta(\omega) + 
\frac{\beta(\bbox{r}_1, \bbox{r}_2,\omega)}{\Delta^2}  
R(\omega) \right] \nonumber \\
& & = - (\pi\nu)^2 \mbox{Re} \langle  Q_{bb}^{11}
(\bbox{ r}_1) Q_{bb}^{22} (\bbox{ r}_2) \rangle_F \nonumber \\
& & - [\mbox{Im}\
G^R(\bbox{ r}_1 - \bbox{ r}_2)]^2 \mbox{Re} \langle Q^{12}_{bb}
(\bbox{ r}_1) Q^{21}_{bb}(\bbox{ r}_1) \rangle_F - (\pi\nu)^2,
\end{eqnarray}
with $G^R$ being the impurity averaged retarded Green's function.
In particular, in the case of 2D and 3D system, $G^R$ is given by
\begin{eqnarray*} 
& & G^R(\bbox{ r}) = \left\{ \begin{array}{lr}
\displaystyle{-i \nu \int_{-\pi/2}^{\pi/2} d\theta \exp [(ip_Fr-r/2l)\cos
\theta ],} & \ 2D\\ 
-\pi\nu(p_F r)^{-1} \exp[ip_Fr
- r/2l], & 3D
\end{array}
\right.
\end{eqnarray*}
where $l$ is the mean free path. The key point making the further
progress possible is that the term containing the single eigenfunction
correlations in lhs of Eq.(\ref{genex}) is proportional to
$\delta(\omega)$, whereas the one depending on the correlations of two
different eigenfunctions is regular at $\omega = 0$. Thus, 
separation of the expression in the
rhs of Eq. (\ref{genex}) into 
the singular (proportional to $\delta(\omega)$) 
and regular parts allows one to obtain the quantities
$\alpha(\bbox{r}_1,\bbox{r}_2)$ and
$\beta(\bbox{r}_1,\bbox{r}_2,\omega)$. 

Now we turn to the case of a metallic system in the weak localization
regime. The corresponding small parameter is given by Eq. (\ref{pi})
below. For further purposes, we introduce the functions
\begin{eqnarray} \label{fun1}
f_1(\bbox{ r}_1, \bbox{ r}_2) & = & \Pi^2 (\bbox{ r}_1,
\bbox{ r}_2), \nonumber \\
f_2(\bbox{ r}_1, \bbox{ r}_2) & = & (2V)^{-1} \int d\bbox{ r}
\left[\Pi^2 (\bbox{ r}, \bbox{ r}_1) + \Pi^2 (\bbox{ r},
\bbox{ r}_2) \right], \nonumber\\
f_3 & = & V^{-2} \int d\bbox{ r}
d\bbox{ r}' \Pi^2 (\bbox{ r}, \bbox{ r}'), \nonumber \\
f_4(\bbox{ r}_1, \bbox{ r}_2) & = & V^{-1} \int d\bbox{ r}
\Pi (\bbox{ r}, \bbox{ r}_1) \Pi (\bbox{ r},
\bbox{ r}_2). 
\end{eqnarray}
Here the diffusion propagator $\Pi$ is the solution to the diffusion
equation
\begin{equation} \label{diff} 
- D \nabla^2\Pi(\bbox{ r}_1,\bbox{ r}_2) = (\pi \nu)^{-1}
\delta(\bbox{ r}_1 - \bbox{ r}_2)   
\end{equation}
with appropriate boundary conditions. We obtain:
\begin{eqnarray} \label{pi1}
\Pi(\bbox{ r}_1,\bbox{ r}_2) & = & (\pi \nu V)^{-1}
\sum_{\bbox{q}} (Dq^2)^{-1} \phi_{\bbox{q}} (\bbox{ r}_1)
\phi_{\bbox{q}} (\bbox{ r}_2), 
\end{eqnarray}
with $\phi_{\bbox{q}}$ being the eigenfunction of the
diffusion operator corresponding to the eigenvalue $Dq^2$, 
$\bbox{q} \ne 0$. 
The level correlation function has the form \cite{KM}
\begin{equation} \label{levcor}
R(\omega) = 1 - s^{-2} \sin^2 s + f_3 \sin^2 s + O(g^{-3})\ , 
\end{equation}
where a dimensionless parameter $s = \pi\omega/\Delta$ is
introduced. The first two terms in Eq. (\ref{levcor}) are given by
RMT, while the third one is the correction of order $g^{-2}$ 
due to the diffusion modes. Here $g = 2\pi E_c/\Delta$ is the
dimensionless conductance, with $E_c$ being the Thouless energy.

The sigma-model correlation functions $\langle Q_{bb}^{11}(\bbox{
r}_1) Q_{bb}^{22} (\bbox{ r}_2) \rangle_F$ and $\langle
Q_{bb}^{12}(\bbox{ r}_1) Q_{bb}^{21} (\bbox{ r}_2) \rangle_F$ can be
calculated for relatively low frequencies $\omega \ll E_c$ with the
use of a general method developed in Refs. \cite{KM,MF} which allows
one to take into account spatial variations of the field $Q$. The
results are obtained in form of expansions in $g^{-1}$. First, we
restrict ourselves to the terms of order $g^{-1}$. Then, the result
for the first correlator reads as  
\begin{eqnarray} \label{1122a}
& & \langle Q_{bb}^{11}(\bbox{ r}_1) Q_{bb}^{22} (\bbox{ r}_2)
\rangle_F \nonumber \\ 
& & \qquad = -1 - 2i {\exp (is) \sin s\over (s+i0)^{2}} - 2i \frac{1}{s+i0}
\Pi(\bbox{ r}_1, \bbox{ r}_2)\ .  
\end{eqnarray}
The first two terms in Eq. (\ref{1122a}) represent the result of
the so-called zero-mode approximation \cite{Efetov}, which takes into
account only the spatially constant configurations of the
field $Q(\bbox{ r})$, so that the
functional integral over $DQ(\bbox{ r})$ is reduced to an integral
over a single matrix $Q$. The last term is the correction of order
$g^{-1}$. An analogous calculation for the second correlator yields 
\cite{GErev}:
\begin{eqnarray} \label{Q3}
&&\langle Q_{bb}^{12} (\bbox{ r}_1) Q_{bb}^{21} (\bbox{ r}_2)
\rangle_F =  -2\left\{ {i\over s+i0 }\right. \nonumber
\\&&\left.\ \ + \left[ 1 +
 i { \exp (is) \sin s\over (s+i0)^{2}} 
\right] \Pi(\bbox{ r}_1, \bbox{ r}_2) \right\}.
\end{eqnarray}

Now, separating regular and singular parts in rhs of Eq. (\ref{genex}),
we obtain the following result for the autocorrelations of the same 
eigenfunction:
\begin{eqnarray} \label{fin1}
& & V^2\langle \vert \psi_k(\bbox{ r}_1) \psi_k(\bbox{ r}_2) \vert^2
\rangle -1  \nonumber \\
& & =   k_d(r) 
 [1 + \Pi(\bbox{ r}_1,\bbox{ r}_1)] + \Pi(\bbox{ r}_1,\bbox{ r}_2),
\end{eqnarray}
and for the correlation of amplitudes of two different eigenfunctions
\begin{equation} \label{fin2}
V^2\langle \vert \psi_k(\bbox{ r}_1) \psi_l(\bbox{ r}_2) \vert^2
\rangle-1  =     k_d(r) 
 \Pi(\bbox{ r}_1,\bbox{ r}_1), \ \ k \ne l 
\end{equation}
Here $r = \vert \bbox{r}_1 - \bbox{r}_2 \vert$, and  
the function $k_d(r)$ is defined as 
\begin{eqnarray*}
& & k_d(r) = (\pi\nu)^{-2}\left[ \mbox{Im} G^R(\bbox{ r})\right]^2 = \\
& & = \exp (-r/l) \left\{
\begin{array}{ll} 
1, & \ \ 2D, \ \ p_Fr \ll 1 \\
8(\pi p_F r)^{-1} \cos^2 p_Fr, & \ \ 2D, \ \ p_Fr \gg 1 \\
(p_Fr)^{-2} \sin^2 p_Fr, & \ \ 3D 
\end{array}
\right.
\end{eqnarray*}
Note that the result (\ref{fin1}) for
$\bbox{ r}_1 = \bbox{ r}_2$ is the inverse participation ratio
previously obtained in Ref. \cite{MF}, while that for arbitrary
spatial separation was found in the zero-mode approximation ($g =
\infty$) in Ref. \cite{Prig}.  

Eq. (\ref{fin2})  shows
that the correlations between different eigenfunctions are relatively
small in the weak disorder regime. Indeed, they are proportional to
the small parameter $\Pi(\bbox{r},\bbox{r})$, which is equal in the
cases of 2D and 3D geometry to ($L$ is the size of the system)
\begin{eqnarray} \label{pi}
\Pi(\bbox{ r}, \bbox{ r}) = \left\{ 
\begin{array}{rlr}
& (\pi g)^{-1} \ln L/l, & \ \ \ 2D\\
\sim & g^{-1} L/l, & \ \ \ 3D
\end{array}
\right.
\end{eqnarray}
In particular, for $\bbox{ r}_1
= \bbox{ r}_2$ we have 
\begin{equation} \label{fin3}
V^2{\langle \vert \psi_k (\bbox{ r}) \psi_l \bbox{ r}) \vert^2
\rangle} - 1 = \delta_{kl} + (1 + \delta_{kl}) \Pi(\bbox{ r},\bbox{ r}).
\end{equation}
The correlations are enhanced by disorder; when the system approaches
the mobility edge in 3D or the strong localization regime in 2D, the
relative magnitude of correlations, $\Pi(\bbox{ r}, \bbox{ r})$
becomes the quantity of order of unity.

An inspection of Eqs. (\ref{fin1}), (\ref{fin2}) shows that while the
correlations of amplitude of the same wavefunction survives for the
large separation between the points, $r \gg l$, and is proportional to
$g^{-1}$, the correlations of two different wavefunctions decay
exponentially for the distances larger than the mean free path
$l$. This is, however, an artifact of the $g^{-1}$ approximation, and
the investigation of the corresponding tails requires the
extension of the above calculation to the terms proportional to
$g^{-2}$. We find that the correlator $\langle Q_{bb}^{11}Q_{bb}^{22}
\rangle_F$ gets the following correction: 
\begin{eqnarray} \label{correct}
& & \delta \langle Q_{bb}^{11}Q_{bb}^{22} \rangle_F = -f_1 + 2f_4 +
\exp(2is) f_3 \nonumber \\
& & \qquad -2i {\exp(2is)\over s+i0} (f_2 - f_3) \nonumber \\
& & \qquad - {\exp(2is)-1\over 2(s+i0)^{2}} (f_1 - 4f_2 + 3f_3 - 4f_4). 
\end{eqnarray}
Consequently, we obtain the following results for the correlations of
different eigenfunctions at $r>l$:
\begin{eqnarray} 
&&V^2\langle \vert \psi_k(\bbox{ r}_1) \psi_l(\bbox{ r}_2) \vert^2
\rangle -1   =  {1\over 2}\left (1 - { \sin^2 s\over s^2}\right)^{-1} 
\nonumber \\
&&\qquad  \times  \left[ f_1 - f_3 - 2f_4 - {2 \sin (2s)\over s} (f_2 - f_3)
\right. \nonumber\\ && \qquad \left.
 - { \sin^2 s\over s^2} (f_1 - 4f_2 + 3f_3 - 2f_4) \right]\ ,\qquad k\ne l 
\label{corr2}
\end{eqnarray}
As it should be expected, the integrals over the coordinate of 
this correlation function are equal to zero. This property is just the
normalization condition and should hold in arbitrary order of
expansion in $g^{-1}$.  Note also that the
correlation of different eigenfunctions becomes now frequency
dependent in a non-trivial way, even in the regime $\omega\ll E_c$
considered.  

The quantities $f_2$, $f_3$, and $f_4$ are proportional to $g^{-2}$,
with some numerical prefactors \cite{KM}. On the other hand, $f_1$ depends
essentially on the distance $r=|\bbox{r}_1-\bbox{r}_2|$. In
particular, for $l\ll r\ll L$ we find
\begin{eqnarray*}
& & f_1(\bbox{r}_1,\bbox{r}_2)=\Pi^2(\bbox{r}_1,\bbox{r}_2)\approx
\left\{
\begin{array}{ll}
\displaystyle{{1\over (\pi g)^2}\ln^2{L\over r}\ }, & \qquad 2D\\
\displaystyle{{1\over (4\pi^2\nu Dr)^2}}        \ , & \qquad 3D
\end{array}
\right.
\end{eqnarray*}
Thus, for $l<r\ll L$, the contributions proportional to $f_1$ dominate
in Eq.(\ref{corr2}), and we get
\begin{equation}
V^2\langle \vert \psi_k(\bbox{ r}_1) \psi_l(\bbox{ r}_2) \vert^2
\rangle -1 = {1\over 2}\Pi^2(\bbox{r}_1,\bbox{r}_2)\ ,\ \ k \ne l.
\label{adm3}
\end{equation}

Another correlation function, generally used for the calculation of
the linear response of the system,
\begin{eqnarray*} 
& & \gamma(\bbox{ r}_1, \bbox{ r}_2, \omega) =
\left\langle \psi^*_k(\bbox{r}_1) \psi_l(\bbox{r}_1) 
\psi_k(\bbox{r}_2) \psi^*_l(\bbox{ r}_2) \right\rangle, \ \ \ k \ne l,
\end{eqnarray*}
can be calculated in a similar way (cf. \cite{GErev}). Starting from
the quantity 
\begin{eqnarray} \label{b1}
& & B(\bbox{ r}_1, \bbox{ r}_2, \omega) = \nonumber \\ &&
\left\langle \sum_{k,l} \psi^*_k(\bbox{r}_1) \psi_l(\bbox{r}_1) 
\psi_k(\bbox{r}_2) \psi^*_l(\bbox{ r}_2)
\delta (\epsilon - \epsilon_k) 
 \delta(\epsilon + \omega
- \epsilon_l) \right\rangle \nonumber\\ && \qquad- \left\langle \sum_{k} 
\psi^*_k(\bbox{r}_1) \psi_k(\bbox{r}_2) \delta(\epsilon - \epsilon_k)
\right\rangle \nonumber \\
& & \qquad \times \left\langle \sum_l
 \psi_l(\bbox{r}_1) \psi^*_l(\bbox{ r}_2) \delta(\epsilon + \omega
- \epsilon_l) \right\rangle \ ,
\end{eqnarray}
and repeating the derivation that led us to Eq. (\ref{genex}), we get
another identity:
\begin{eqnarray} \label{genex1}
& & 2\pi^2 \left[ \frac{\alpha(\bbox{r}_1, \bbox{r}_2)}{\Delta}
\delta(\omega) + 
\frac{\gamma(\bbox{r}_1, \bbox{r}_2,\omega)}{\Delta^2}  
R(\omega) \right] \nonumber \\
& & = - (\pi\nu)^2 \mbox{Re} \left\{ \langle  Q_{bb}^{12}
(\bbox{ r}_1) Q_{bb}^{21} (\bbox{ r}_2) \rangle_F \right. \nonumber \\
& & \left. - k_d(r) \langle Q^{11}_{bb}
(\bbox{ r}_1) Q^{22}_{bb}(\bbox{ r}_1) \rangle_F - k_d(r) \right\},
\end{eqnarray}
Taking into account Eqs. (\ref{1122a}) and (\ref{Q3}), and separating
the rhs into the regular and singular parts, we recover
Eq. (\ref{fin1}) and obtain
\begin{eqnarray} \label{states}
& & V^2 \langle \psi^*_k(\bbox{r}_1) \psi_l(\bbox{r}_1) 
\psi_k(\bbox{r}_2) \psi^*_l(\bbox{ r}_2) \rangle \nonumber \\
& & \qquad = k_d(r) + \Pi(\bbox{
r}_1,\bbox{ r}_2),\ \ \ k \ne l. 
\end{eqnarray}

As was mentioned, the above derivation is valid for $\omega \ll
E_c$. In order to obtain the results in the range $\omega \gtrsim E_c$ one
can calculate the sigma-model correlation functions entering
Eqs. (\ref{genex}), (\ref{genex1}) by means of the perturbation
theory \cite{ASh}. We find then for $k \ne l$
\begin{eqnarray} \label{pert}
& & V^2\langle \vert \psi_k(\bbox{ r}_1) \psi_l(\bbox{ r}_2) \vert^2
\rangle = 1 + \mbox{Re} \left\{ k_d(r) \Pi_{\omega}(\bbox{r}_1,
\bbox{r}_2) \right. \nonumber \\
& & \left. \qquad + \frac{1}{2} \left[ \Pi^2_{\omega} (\bbox{r}_1,\bbox{r}_2)
- \frac{1}{V^2} \int d\bbox{r} d\bbox{r}' \Pi^2_{\omega}
(\bbox{r},\bbox{r}') \right] \right\}, \\
& & V^2 \langle \psi^*_k(\bbox{r}_1) \psi_l(\bbox{r}_1) 
\psi_k(\bbox{r}_2) \psi^*_l(\bbox{ r}_2) \rangle = k_d(r) + \mbox{Re}
\Pi_{\omega} (\bbox{r}_1,\bbox{r}_2), \nonumber
\end{eqnarray}
where $\Pi_{\omega} (\bbox{r}_1,\bbox{r}_2)$ is the finite-frequency
diffusion propagator
\begin{equation} \label{Ask}
\Pi_{\omega} (\bbox{r}_1,\bbox{r}_2) = (\pi\nu V)^{-1} \sum_{\bbox{q}}
\frac{ \phi_q (\bbox{r}_1) \phi_q (\bbox{r}_2)}{Dq^2 - i\omega}, 
\end{equation}
and the summation in Eq. (\ref{Ask}) now includes $\bbox{q} = 0$.

The above derivation was performed for the case of a disordered
sample. However, one can repeat it for a classically chaotic ballistic
system, using the recent derivation of the supersymmetric
sigma-model for this case \cite{ball}. In the end one gets the
 analogous results, but with the diffusion
operator being replaced by the Perron-Frobenius one for the given
chaotic system. To analyze the results quantitatively in this case,
one needs then an information about the eigenfunctions and eigenvalues
of the Perron-Frobenius operator. 
This question deserves further investigation and goes
beyond the scope of this Letter.

Finally, we discuss the relation between our results and the recent
experiment \cite{Folk}, where strong correlations in amplitudes of
neighboring conductance peaks of a quantum dot were observed. Within
the one-electron picture, this seems to contradict our result
(\ref{fin2}) of weakness of correlations of different
eigenfunctions. In principle, one could imagine that the dot was far from
the universal (RMT) regime, so that the parameter
$\Pi(\bbox{r},\bbox{r})$ determining the magnitude of correlations was
not small. This would be however in contradiction with the fact that
the total distribution of peak heights in Ref.\cite{Folk} was well
described by the RMT formulae \cite{Jal}, since the corrections to the
distribution of $|\psi^2(\bbox{r})|$ (and consequently to that of peak
heights) are proportional to the same parameter
$\Pi(\bbox{r},\bbox{r})$ \cite{MF}. Another possibility is that the
electron-electron interaction effects lead to some modification of
eigenfunction correlations. This problem remains to be studied in future.

We are grateful to Charlie Marcus for a discussion which
stimulated this research. The work was supported by the Alexander von
Humboldt Foundation (Y.~M.~B.) and SFB195 der Deutschen
Forschungsgemeinschaft (A.~D.~M.).

\end{document}